\documentclass[12pt]{article}
\usepackage{amssymb}
\usepackage{amsfonts}
\usepackage{amsmath,amsthm}
\usepackage{subfigure}
\usepackage{graphicx,epsfig,color }

\oddsidemargin 10mm \numberwithin{equation}{section}
\def\be{\begin{equation}}
\def\ee{\end{equation}}
\usepackage[numbers,sort&compress]{natbib}
\usepackage[top=3.5cm,right=3cm,bottom=3cm,left=2cm]{geometry}
\begin{document}
\begin{center}
{{\bf {Central AdS Generalized Ay{\'o}n-Beato-Garc{\'\i}a Black
Holes and Joule-Thomson Expansion}} }
 \vskip 1 cm {Elham Ghasemi
\footnote{E-mail address: e\_ ghasemi@semnan.ac.ir} and Hossein
Ghaffarnejad \footnote{E-mail address:
hghafarnejad@semnan.ac.ir; ORCID: 0000-0002-0438-6452}}\\
\vskip 0.1 cm
{\textit{Faculty of Physics, Semnan University, P.C. 35131-19111, Semnan, Iran} } \\
\end{center}
\begin{abstract}
In this paper, the Joule-Thomson (JT) adiabatic expansion is
investigated for generalized Ayon-Beato-Garcia (ABG) regular black
hole. It has a magnetic charge which makes central region of the
black hole metric  to be AdS spacetime and so become non singular.
Thus we not need to use an additional cosmological parameter
coming from ADS/CFT correspondence for production of pressure
coordinate in the black hole equation of state. Form this point of
view our work is a new approach versus to conventional methods
which are addressed in the references of this paper. However we
will see important behavior of parameters of this modified ABG
black hole on possibility of being of JT adiabatic expansion. This
black hole is characterized by five parameters including mass $m$,
magnetic charge $q$, and three other parameters related to form of
used nonlinear electromagnetic interaction fields. Inversion
points are in fact a particular T-P curve at constant enthalpy and
it separates cooling and heating phase of the modified AdS ABG
black hole.
\end{abstract}
\section{Introduction}
Due to the similar behavior of black holes to the ordinary
thermodynamic systems \cite{0}, it is possible to attribute
properties of the ordinary thermodynamic systems to the black
holes. Clausius-Clapeyron relation, Joule-Thomson expansion, the
heat engines subject, Maxwell`s equal-area law, and the chemical
behavior are some examples which have been investigated for AdS
charged black holes in the literature \cite{1,2,3,4,5}. Also, the
mentioned subjects have been studied for AdS-Kerr black holes in
the same way \cite{6,7,8,9,10}. Moreover, these topics were
studied for other black holes in the alternative theories of
gravity \cite{11,12,13,14,15,16,17}. In all of them the
cosmological constant plays as thermodynamic pressure in the black
hole system which is called as extended phase space. Fortunately
central region of the ABG black hole behaves asymptotically as AdS
space in which its magnetic charge produces a variable
cosmological parameter or variable pressure. This can be
considered as alternative AdS pressure instead of auxiliary
cosmological constant which come from AdS/CFT correspondence and
it is used in usual way \cite{20}. In the present work, we
specifically investigate the JT expansion for a generalized ABG
black hole which is introduced by Cai and Miao \cite{18}. Firstly,
the ABG black holes was proposed by Ay{\'o}n-Beato and Garc{\'\i}a
\cite{19} as a black hole without a central singularity. This
advantage is due to the presence of nonlinear interaction of
electromagnetic fields. The generalized ABG black hole is defined
in terms of five parameters including mass $m$, magnetic charge
$q$, and three other constant parameters which is related to kind
of nonlinear interaction of the electromagnetic fields. In other
words these three parameters can be control regularity of the
central region of the ABG black hole. Thermodynamics and phase
transition of the mentioned black hole is studied in our recently
work \cite{20} and we investigate effects of the parameters of
this kind of regular black hole on the JT adiabatic
expansion now. \\
As we said above, in fact, the study of the JT phenomenon in black
holes is modeled from the behavior of real gases, which is studied
in ordinary thermodynamics. As the small-to-large scales phase
transition in black holes corresponds to the liquid-to-gas phase
transition in the Van der Waals fluid, because the behavior of
temperature and pressure changes is the same for both. Of course,
for this phase transition to happen, conditions must be
established which in black holes, the most important quantity is
the pressure of the surrounding environment on the surface of the
black hole's horizon and it is provided by the negative
cosmological constant. It is because of AdS/CFT correspondence .
In the case of JT adiabatic expansion, we follow similar approach
too.  Generally, if a gas become cooler (warmer) when we raise the
affecting pressure, then its temperature decreases (increases) at
constant entropy. This behavior for gases is called JT adiabatic
expansion and is happened some times. Simplest tools to check that
do it happened for a particular gas is the `inversion
temperature-pressure curve`. In fact it determines some points on
the T-P phase space at constant enthalpy such that the gas under
pressure can may be participates in the JT phenomena. This pattern
is also followed in black holes. As an example we like to
investigate this effect for modified AdS ABG black hole here.
 Since a black hole's mass plays
the role of enthalpy in the extended phase space, the JT expansion
could be examined for black holes as well. In this way, the black
hole's temperature changes as a result of its pressure variation
while its mass remains constant. The temperature variation due to
the pressure variation is exhibited as isenthalpic curves and it
is introduced by a JT coefficient  $\mu=\left( \frac{\partial
T}{\partial P}\right)_m.$ This is in fact slope of T-P diagrams at
constant enthalpy. Hence one can infer that the sign of the JT
coefficient determines the behavior of the black hole as a heater
or cooler. In this view when $\mu>0 (\mu<0)$ then the black hole
is heater (cooler) because of increase (decrease) of its
temperature versus the raise of AdS pressure. So it can be
concluded that it is located in the heating (cooling) phase.
Inversion temperature and inversion pressure are another factors
are studying in this way. The inversion point $(T_i, P_i)$ is
obtained from $\mu=0$ which is the maximum point of the
isenthalpic curves. In fact, the inversion curve as a separation
line between two heating and cooling phases passes through the
inversion points of the different isenthalpic curves. In other
words the intersection of the inversion curves and isenthalpic
curves are positions where the black hole participates in the
cooling-heating phase transition. One can follows
\cite{21,22,23,24,25,26,27,28,29,30, 31,32,33,34,35,36,37} where
the JT expansion are examined for various types of black holes.\\
At last, in general, only when a black hole participates in the
Joule-Thomson phase transition does the inversion curve intersect
with the temperature-pressure isenthalpic curves, and this
possibility depends on the parameters and characteristics of the
black hole system. The layout of this work is as follows.\\
 In section 2 we give a short review about the generalized ABG black hole defined by \cite{18} and in
section 3 we generate some suitable equations related to the JT
coefficient. In this section we plot inversion curves and
isenthalpic diagrams for different values of the black hole
parameters. Last section denotes to concluding remark.
\section{Generalized ABG black hole and its thermodynamic}
Let us start with line element of the generalized ABG magnetic
spherically symmetric static black hole which in the Schwarzschild
frame is defined by \cite{18}:
\begin{equation}
ds^2=-f(r)dt^2+f(r)^{-1}dr^2+r^2(d\theta^2+sin\theta^2d\varphi ^2)
\end{equation}
in which the metric potential has the following form.
\begin{equation}\label{22}
f(r)=1-\frac{2mr^{\frac{\alpha\gamma}{2}-1}}{(q^\gamma+r^\gamma)^{\alpha/2}}
+\frac{q^2r^{\frac{\beta\gamma}{2}-2}}{(q^\gamma+r^\gamma)^{\beta/2}}
\end{equation}
where $m$ and $q$ refer to the mass and magnetic charge of the
generalized ABG black hole. The three parameters of $\alpha$,
$\beta$, and $\gamma$ are related to each other through
$\alpha\gamma\geq6$, $\beta\gamma\geq8$ and $\gamma>0$ \cite{18}.
In fact regularity of this black hole depends to these
inequalities between the parameters. If we choose different values
for these three parameters by regarding the above inequalities we
can obtain different kinds of regular modified ABG black holes. In
this paper we use ansatz $\alpha\gamma=6$ and $\alpha\beta=8$ for
which the metric potential (\ref{22})  takes on simpler form given
in the paper \cite{20}. In that work we studied  thermodynamic
phase transition of the modified AdS ABG black hole. Also we
showed that the above metric can be viewed as a Schwarzschild-de
Sitter form of the black hole metric if we set a variable mass
function and the variable cosmological parameter as follows.
\begin{equation}M(r)=\frac{mr^\frac{\alpha\beta}{2}}{(r^\gamma+q^\gamma)^\frac{\alpha}{2}},~~~\Lambda(r)=-\frac{3q^2r^{\frac{\beta\gamma}{2}-4}}{(r^\gamma+q^\gamma)^\frac{\beta}{2}}
.\end{equation} Regarding the AdS space pressure definition
$P=\frac{-\Lambda}{8\pi}$ for this variable cosmological parameter
and by substituting it into the event horizon equation $f(r_+)=0$
we can obtain pressure equation versus the black hole mass $m$ and
the other parameters such that
\begin{equation}\label{P}
P(m,r_+)=\frac{6mr_+^{\frac{\alpha\gamma}{2}-1}-3(q^\gamma+r_+^\gamma)^{\alpha/2}}
{8\pi r_+^2(q^\gamma+r_+^\gamma)^{\alpha/2}}.
\end{equation}
This form of the state equation is suitable to study isenthalpic
phenomena because the black hole mass $m$ plays the enthalpy role,
so it remains constant during the JT process. Moreover, the
Hawking temperature is obtained through the black hole surface
gravity on the horizon, $T_H=\frac{\kappa}{2\pi}|_{r=r_+}$
\cite{20} such that
$$T(m,r_+)=\frac{mr_+^{\frac{\alpha\gamma}{2}-2}}{4\pi (q^\gamma+r_+^\gamma)^{\alpha/2}}
\left[\frac{(2-\alpha\gamma)(q^\gamma+r_+^\gamma)+\alpha\gamma r_+^\gamma}
{(q^\gamma+r_+^\gamma)} \right] $$
\begin{equation}\label{T}
+\frac{q^2r_+^{\frac{\beta\gamma}{2}-3}}{8\pi (q^\gamma+r_+^\gamma)^{\beta/2}}
\left[\frac{(\beta\gamma-4)(q^\gamma+r_+^\gamma)-\beta\gamma r_+^\gamma}
{(q^\gamma+r_+^\gamma)} \right]
\end{equation}
Considering the dimensionless parameter of $x=\frac{r_+}{q}$,
relations of \eqref{P} and \eqref{T} are rewritten as:
\begin{equation}\label{P1}
P(m,x)=\frac{6mx^{\frac{\alpha\gamma}{2}-1}-3q(1+x^\gamma)^{\alpha/2}}
{8\pi q^3x^2(1+x^\gamma)^{\alpha/2}}
\end{equation}
and
$$T(m,x)=\frac{mx^{\frac{\alpha\gamma}{2}-2}}{4\pi q^2(1+x^\gamma)^{\alpha/2}}
\left[\frac{(2-\alpha\gamma)(1+x^\gamma)+\alpha\gamma x^\gamma}
{(1+x^\gamma)} \right] $$
\begin{equation}\label{T1}
+\frac{x^{\frac{\beta\gamma}{2}-3}}{8\pi q(1+x^\gamma)^{\beta/2}}
\left[\frac{(\beta\gamma-4)(1+x^\gamma)-\beta\gamma x^\gamma}
{(1+x^\gamma)} \right].
\end{equation}
We are now in position to calculate JT coefficient and investigate
cooling-heating phase transition of this kind of black hole.
\section{Joule-Thomson Expansion}
JT expansion in ordinary thermodynamics is known as an
irreversible process, where a gas moves from a region with high
pressure to a region with low pressure through a porous plug.
Investigating the temperature changes during this expansion is the
main aim of this study. Since the system's enthalpy remains
constant during this process, a set of (T,P) values forms the
isenthalpic curves, where JT coefficient defined as the
isenthalpic curve slope and is given as follows:
\begin{equation}
\mu _{JT}=\left( \frac{\partial T}{\partial P}\right) _H.
\end{equation}
During the JT expansion, the pressure always reduces. So, the sign
of the JT coefficient determines the heating (the temperature
increase) or cooling (the temperature decrease) behavior of the
system. For this purpose, the JT coefficient for a black hole
under consideration is obtained as follow:
\begin{equation}\label{mu}
\mu(m,x)=\left( \frac{\partial T}{\partial P}\right) _m=\left( \frac{\partial T}{\partial x}\right) _m
\left( \frac{\partial x}{\partial P}\right) _m
\end{equation}
As we said in the previous section to choose a particular form of
the modified regular AdS ABG black hole metric given in the ref
\cite{20} we substitute $\alpha=\frac{3\gamma}{4}$ and
$\beta=\frac{8}{\gamma}$  into the relations of (\ref{P1}) and
(\ref{T1}), then we calculate (\ref{mu}) by keeping $m$ as
constant enthalpy such that
$$\mu(m,x)= \Bigg\{mq x^{\frac{3\gamma^2}{8}}(1+x^\gamma)^\frac{4}{\gamma}
\left[-128 x^{2\gamma}
+8(3\gamma^3+9\gamma^2-32)x^\gamma+(-9\gamma^4+72\gamma^2-128) \right] $$
$$+64q^2x^3 (1+x^\gamma)^{\frac{3\gamma}{8}}
\left[+3 x^{2\gamma}-2(\gamma+2)x^\gamma +1\right] \Bigg\} \Bigg/$$
\begin{equation}
\Bigg\{12(1+x^\gamma)^{\frac{4}{\gamma}+1}
\left[3mx^{\frac{3\gamma^2}{8}-1}\left(\gamma^2-8x^\gamma-8
\right) +8q(1+x^\gamma)^{\frac{3\gamma}{8}+1}\right] \Bigg\}.
\end{equation}
To have inversion curves we should obtain roots of the equation
$\mu(m,x)=0$ which reads to the following parametric identity.
$$m_i(x)=\Bigg\{ 64qx(1+x^\gamma)^{\frac{3\gamma}{8}}
\left[ 3x^{2\gamma}-2(\gamma+2)x^\gamma+1\right]
\Bigg\} \Bigg/$$
\begin{equation}
\Bigg\{x^{\frac{3\gamma^2}{8}-2}(1+x^\gamma)^{4/\gamma}
\left[ 128x^{2\gamma}-8(3\gamma^3+9\gamma^2-32)x^\gamma
+9\gamma^4-72\gamma^2+128\right] \Bigg\}.
\end{equation}
In this step, the inversion pressure $P_i$ and inversion temperature $T_i$ are accessible
by substituting $m_i$ into \eqref{P1} and \eqref{T1} as follows:
$$P_i(x)=\Big\{ \left[1152x^{2+2\gamma}-(768\gamma+1536)x^{2+\gamma}+384x^2
\right] (1+x^\gamma)^{-4/\gamma}$$
$$-384x^{2\gamma}+(72\gamma^3+216\gamma^2-768)x^\gamma
-27\gamma^4+216\gamma^2-384\Big\} \Big/$$
\begin{equation}
\Big\{ 8\pi q^2 x^2 \left[128x^{2\gamma}-(24\gamma^3+72\gamma^2-256)x^\gamma
+9\gamma^4-72\gamma^2+128\right] \Big\}
\end{equation}
and
$$T_i(x)=\Big\{x(1+x^\gamma)^{\frac{-(4+\gamma)}{\gamma}}
\big[64x^{3\gamma}+(24\gamma^3-128\gamma-192)x^{2\gamma}$$
$$-(9\gamma^4-24\gamma^3-96\gamma^2+128\gamma+64)x^\gamma
+9\gamma^4-96\gamma^2+192\big] \Big\} \Big/$$
\begin{equation}
\Big\{2\pi q\left[
128x^{2\gamma}-(24\gamma^3+72\gamma^2-256)x^\gamma
+9\gamma^4-72\gamma^2+128\right]\Big\}.
\end{equation}
Now, it is possible to plot the inversion curve in the T-P diagram
for the black hole under consideration  by choosing different
values for $\gamma$ and $q$ (see black colored dash lines in
figures 1,2,3,4,5,6,7,8). In fact, the inversion curve is the
separation boundary between the heating and cooling areas of the
black hole which intersects the isenthalpic curves at the maximum
points given by $\mu=0$ or the inversion points $(T_i, P_i)$. By
looking at these diagrams one can infer that the top of the
isenthalpic curves versus to the inversion curve ($\mu>0$)
represents the heating phase where the temperature increases by
raising the pressure while the below part of these diagrams with
respect to the inversion curve $\mu<0$ shows that the black hole
participates in the cooling phase where the temperature decreases
by raising the pressure. In this paper, we plotted T-P diagrams at
constant enthalpy (the black hole mass) for various values of the
black hole mass parameter ($m$) by considering the positive values
for $\gamma$ which comes from regularity condition of the black
hole metric given in the \cite{18} and we mentioned in the
previous
 section,  but both positive and negative signs for the magnetic charge $q$.\\
The isenthalpic T-P curves plotted in figures 1, 4 and 7 show that
for $\gamma=3$ there is just one inversion curve for $q<0$ while
two inversion curves for $q>0.$ In other choices for fixed
$\gamma=10,20$ there are obtained two inversion curves for both
positive and negative magnetic charges. Looking at the diagrams
one can infer that for a fixed $\gamma$ the JT expansion appears
at lower pressures by raising the absolute value of the magnetic
charge $|q|.$ All figures show that there is not appear the JT
expansion or cooling/heating phase transition for  the lightest
black hole (with smallest mass). In these diagrams intersection of
inversion curves (the black dash lines) and isenthalpic curves
(colored solid lines) namely the maximum point of the isenthalpic
curves, show some positions where the JT expansion are happened.
In other words for some (smallest black holes) which there is no
any intersection point between the isenthalpic curves and the
inversion curve one can result that the central AdS modified ABG
black hole does not participate in the JT adiabatic expansion.
With this regard we obtain that smallest black holes can not
expand adiabatically and remain in thermal equilibrium with its
surroundings. We should point that when mass of an evaporating
quantum black hole remain as constant and so the entropy remains
constant too. In this case the black hole maybe participates in
the JT adiabatic expansion or may not be. Such a black hole
exchanges heat with its surroundings if it participates in the
Joule-Thomson expansion process, otherwise it will be in thermal
equilibrium.
\section{Concluding remark}
In this work we use central AdS generalized ABG magnetically
charged black hole with multiple parameters and studied
cooling/heating phase transition for it. This is done by studying
the isenthalpic T-P curves intersecting with inversion curves.
Inversion curves are obtained when the JT coefficient takes on
zero value. Our mathematical calculation predicted that for
constant magnetic charges the JT expansion appears for massive
generalized ABG black hole by raising the $\gamma$ parameter of
the black hole and low mass black holes do not participate in the
JT expansion phenomena. Because there is not a intersection point
between the isenthalpic curves of the lightest generalized ABG
black hole and corresponding inversion curve. In other words
smallest scale of this kind of black holes remain equilibrium
thermodynamically with its surroundings.  Also for a fixed
$\gamma$ parameter the JT expansion appears for massive
generalized ABG black holes at lower pressures by raising the
magnetic charge parameter. As an future work about the generalized
ABG black hole we like to study effect of holographic entanglement
entropy and its thermalization affects.

\begin{figure}[ht]
\hspace{3mm} \subfigure[{}]{\label{a}
\includegraphics[width=0.45\textwidth]{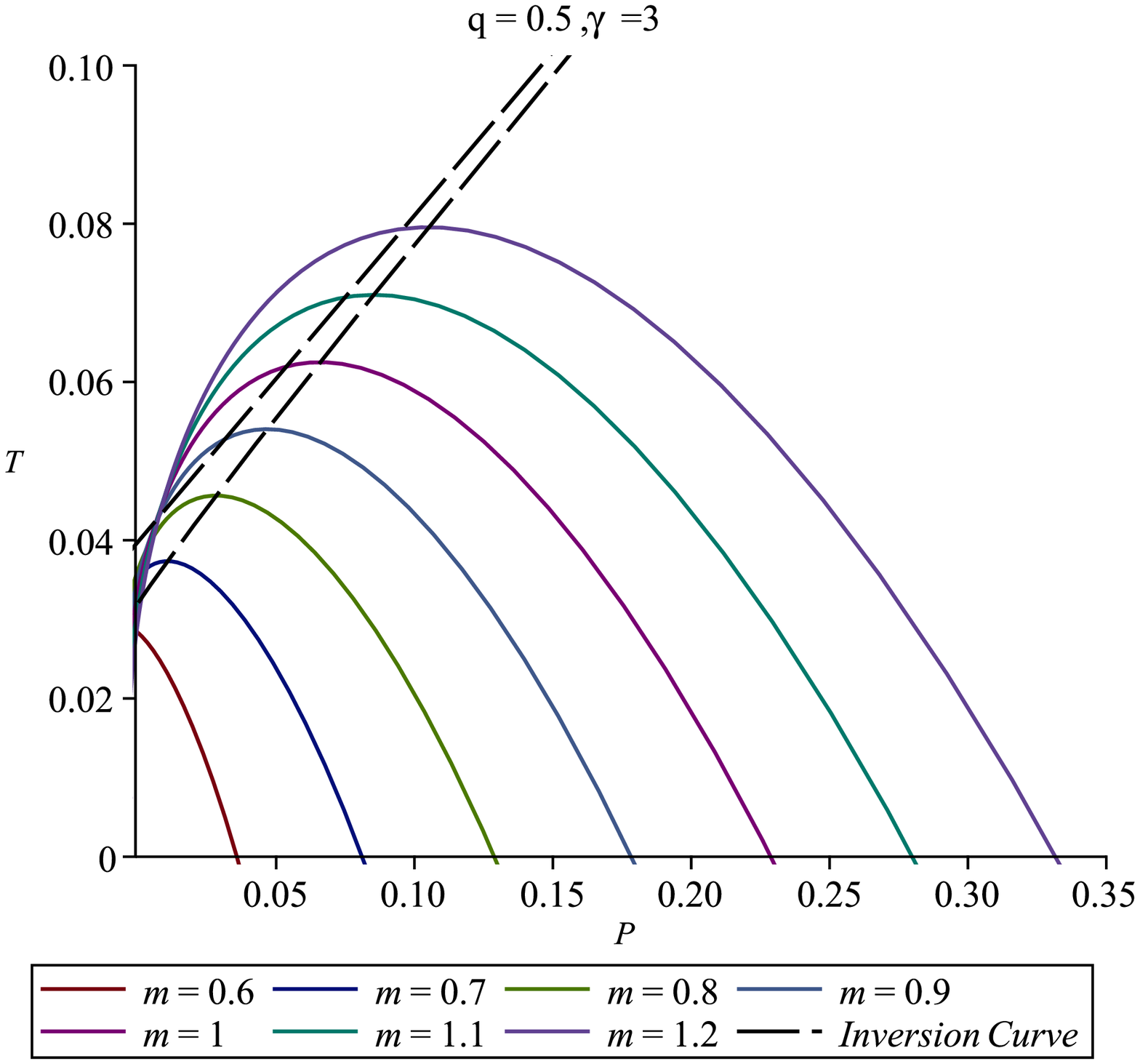}}
\hspace{3mm} \subfigure[{}]{\label{b}
\includegraphics[width=0.45\textwidth]{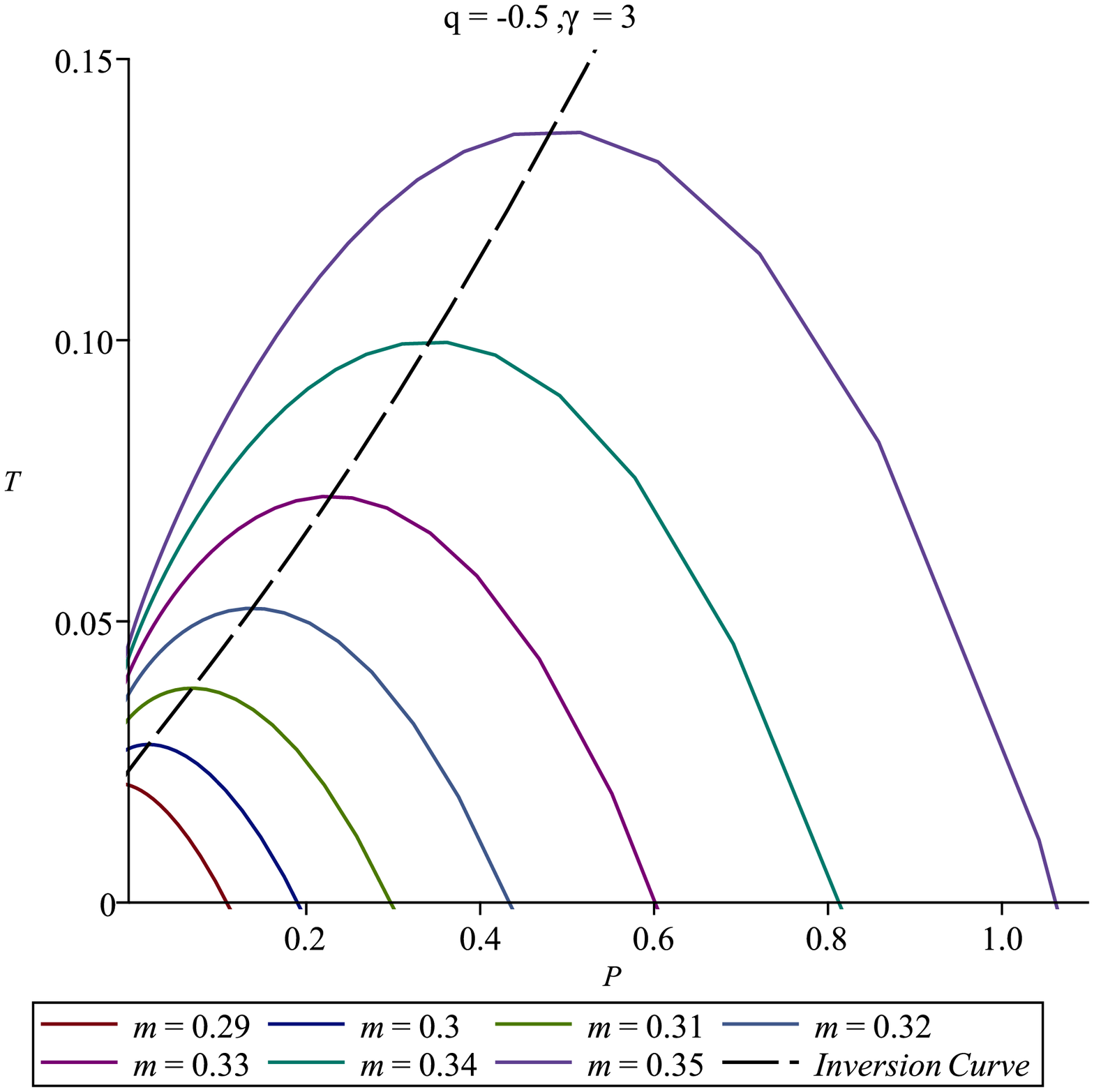}}
\hspace{3mm} \caption{\footnotesize{Intersection of the inversion
curve (the black line) with isenthalpic curves for $q=\pm 0.5$ and
$\gamma=3$}}
\end{figure}
\begin{figure}[ht]
\hspace{3mm} \subfigure[{}]{\label{b1}
\includegraphics[width=0.45\textwidth]{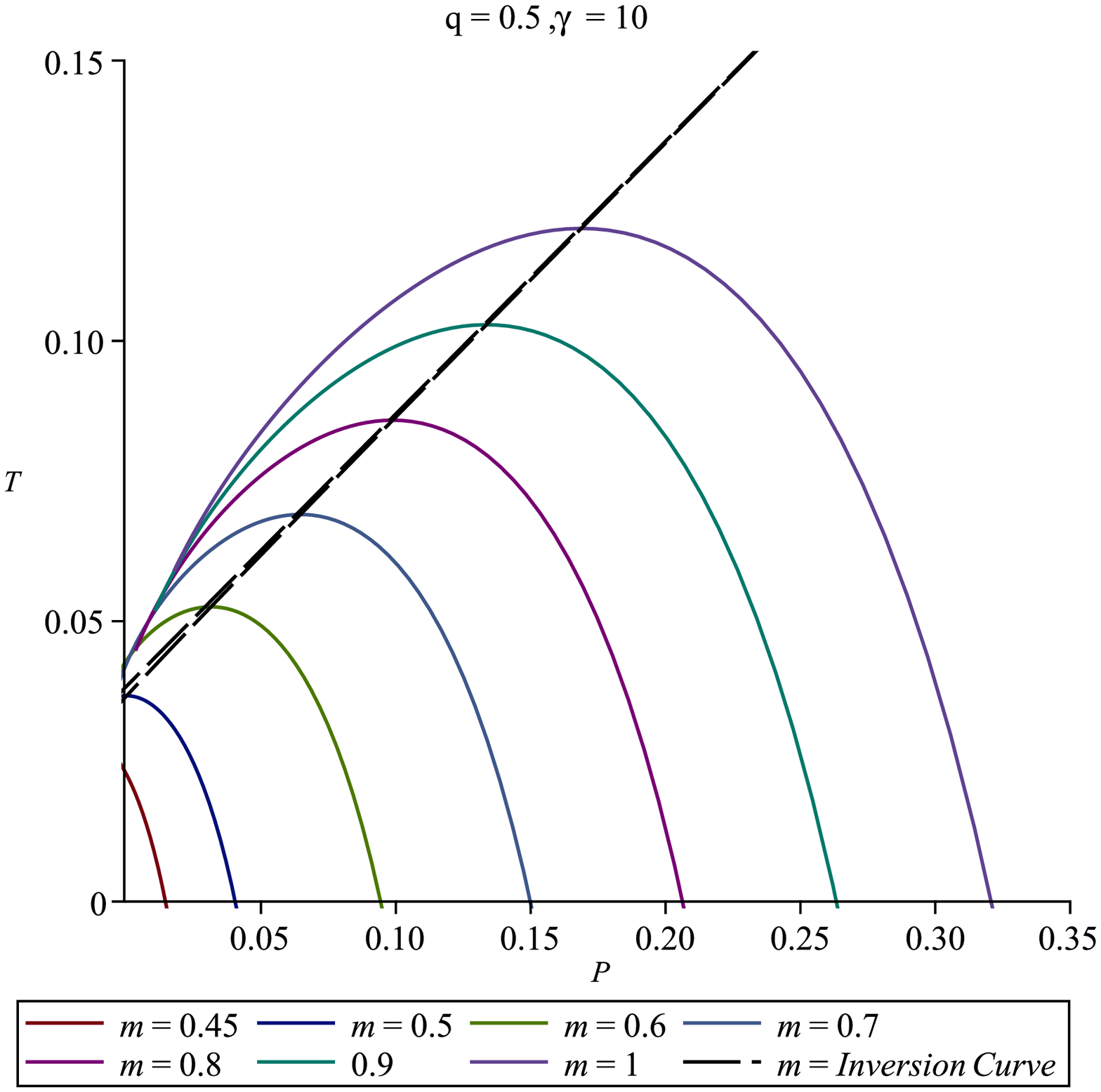}}
\hspace{3mm} \subfigure[{}]{\label{d}
\includegraphics[width=0.45\textwidth]{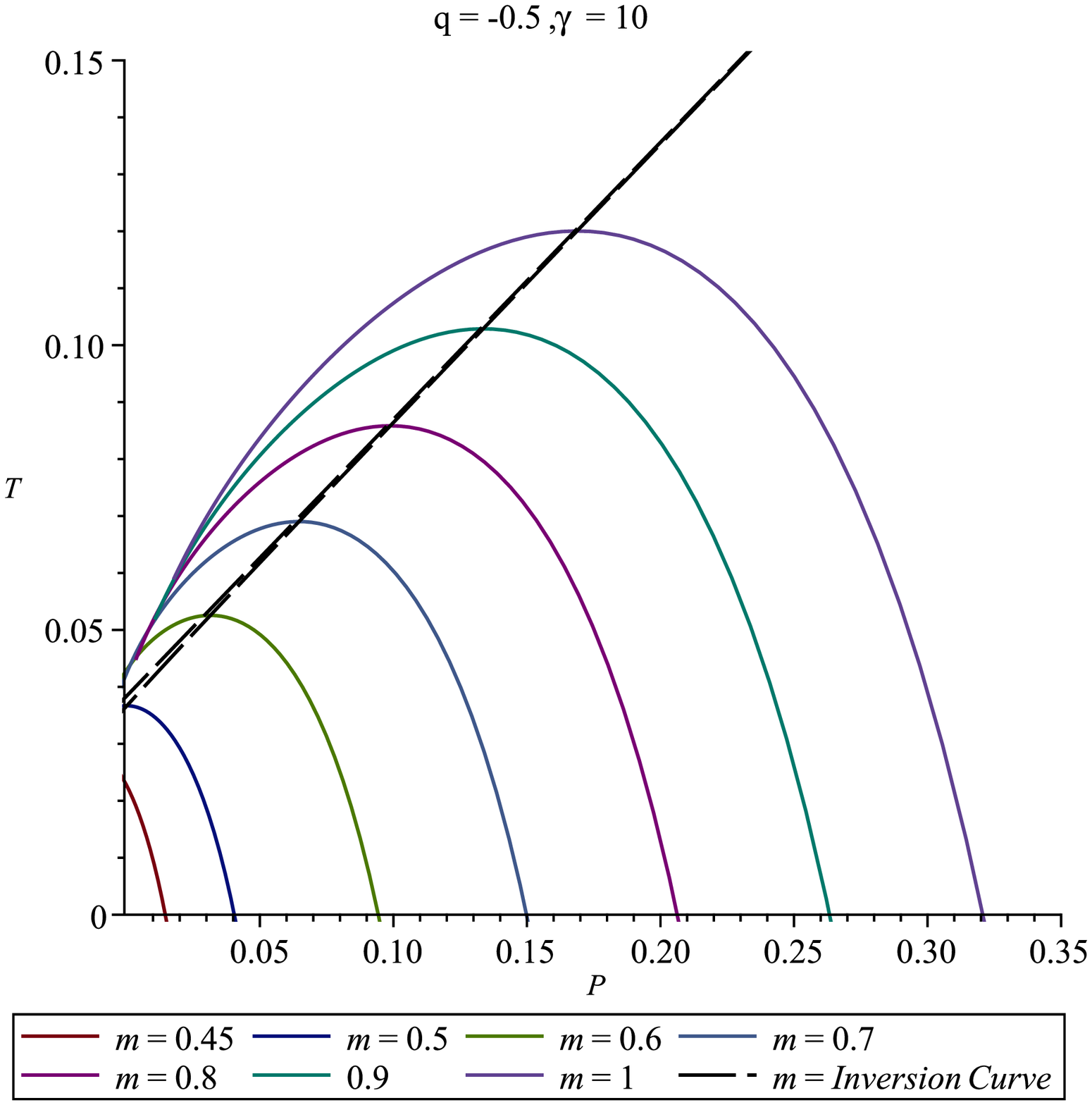}}
\hspace{3mm} \caption{\footnotesize{Intersection of the inversion
curve (the black line) with isenthalpic curves for $q=\pm 0.5$ and
$\gamma=10$}}
\end{figure}
\begin{figure}[ht]
\hspace{3mm} \subfigure[{}]{\label{e}
\includegraphics[width=0.45\textwidth]{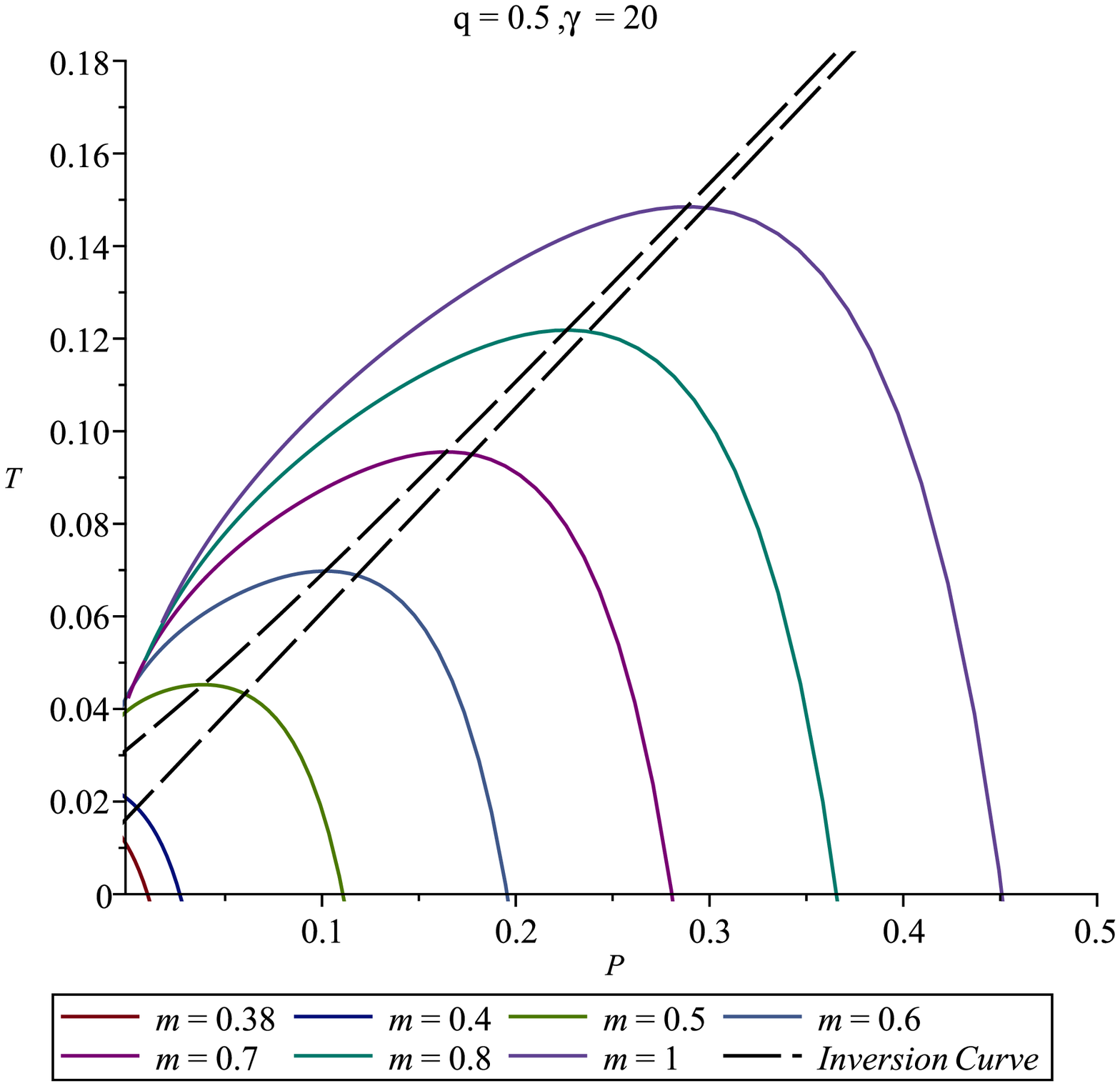}}
\hspace{3mm} \subfigure[{}]{\label{f}
\includegraphics[width=0.45\textwidth]{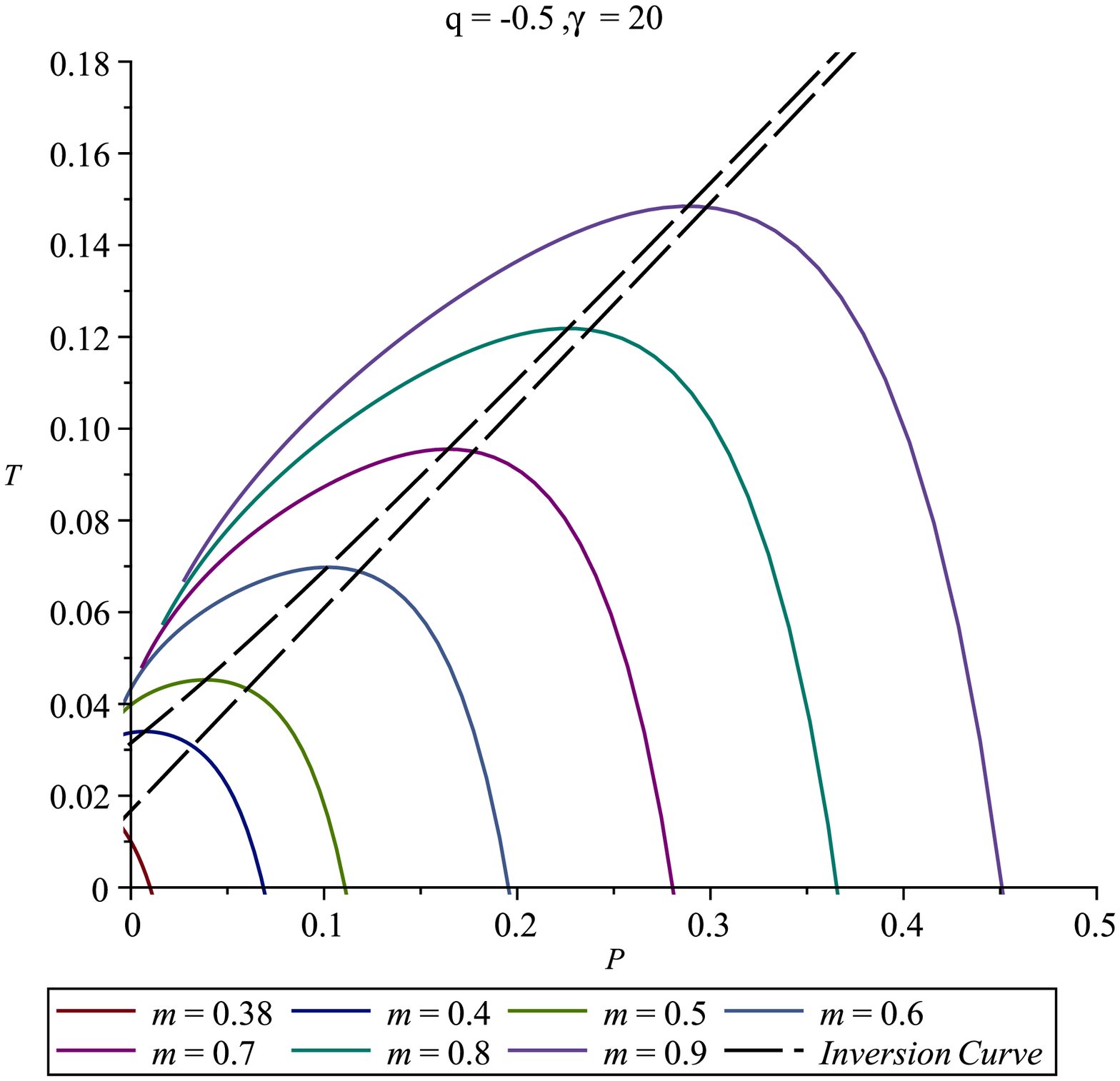}}
\hspace{3mm} \caption{\footnotesize{Intersection of the inversion
curve (the black line) with isenthalpic curves for $q=\pm0.5$ and
$\gamma=20$}}
\end{figure}
\begin{figure}[ht]
\hspace{3mm} \subfigure[{}]{\label{aa}
\includegraphics[width=0.45\textwidth]{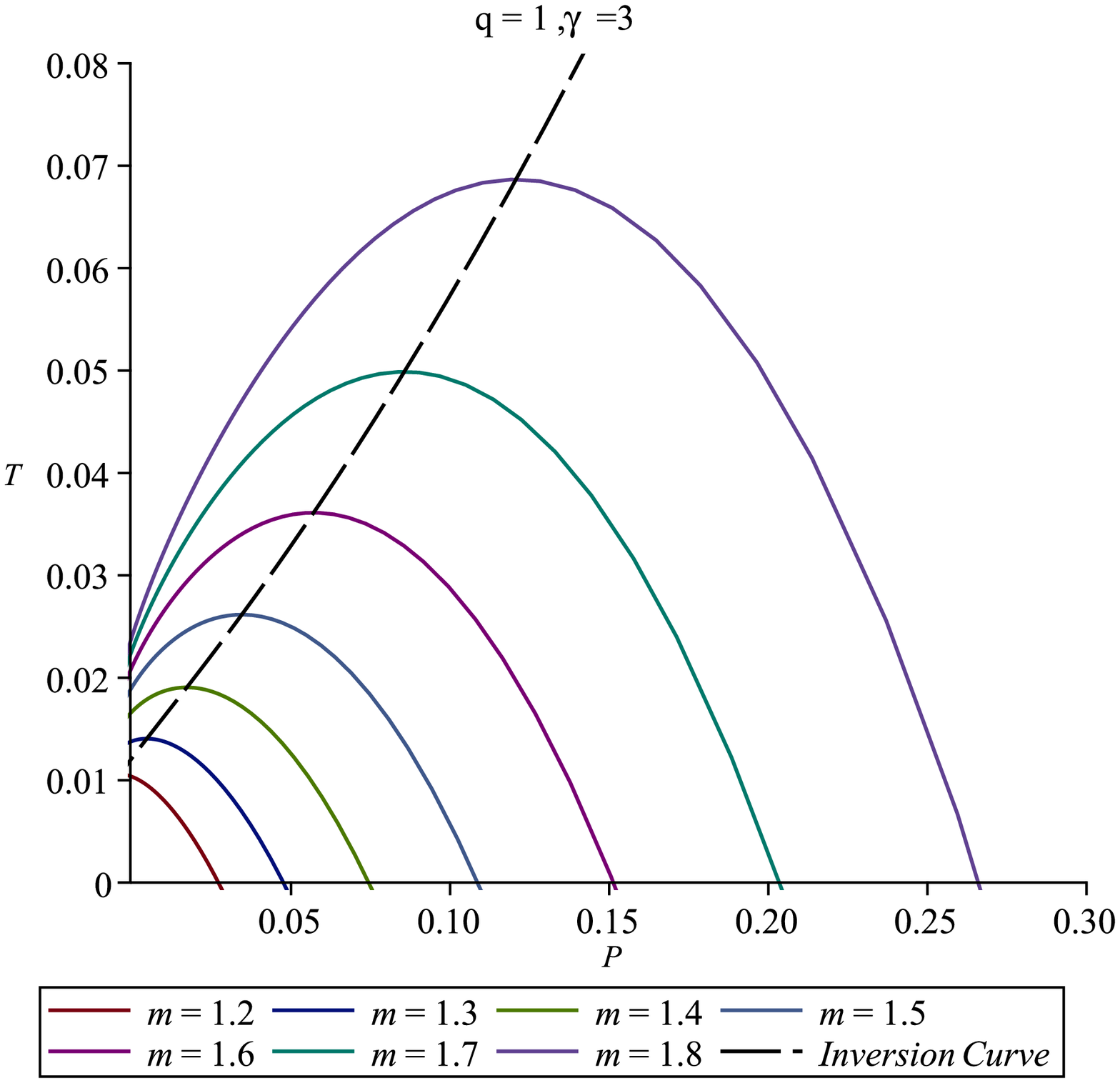}}
\hspace{3mm} \subfigure[{}]{\label{bb}
\includegraphics[width=0.45\textwidth]{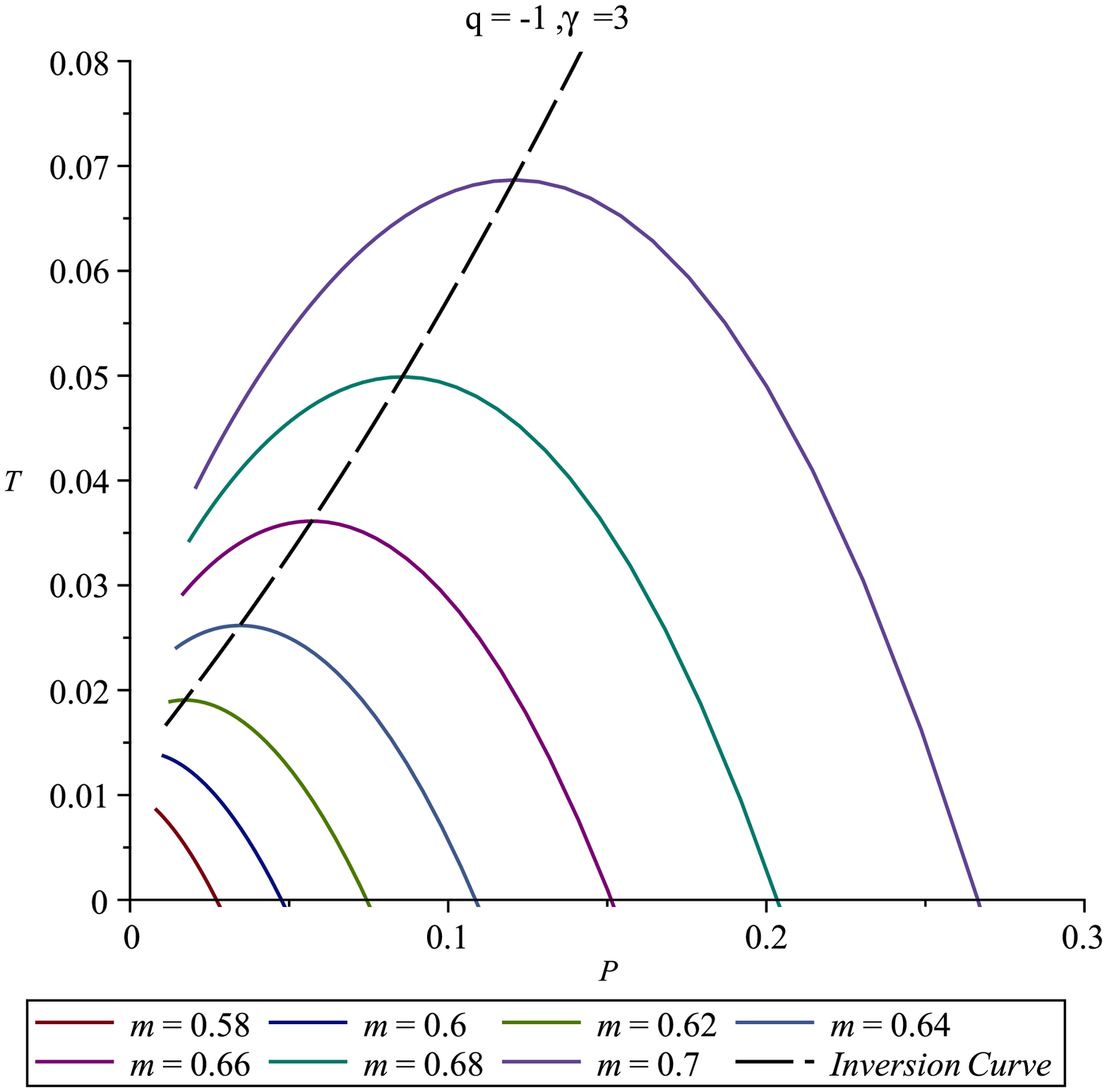}}
\hspace{3mm} \caption{\footnotesize{Intersection of the inversion
curve (the black line) with isenthalpic curves for $q=\pm 1$ and
$\gamma=3$}}
\end{figure}
\begin{figure}[ht]
\hspace{3mm} \subfigure[{}]{\label{cc}
\includegraphics[width=0.45\textwidth]{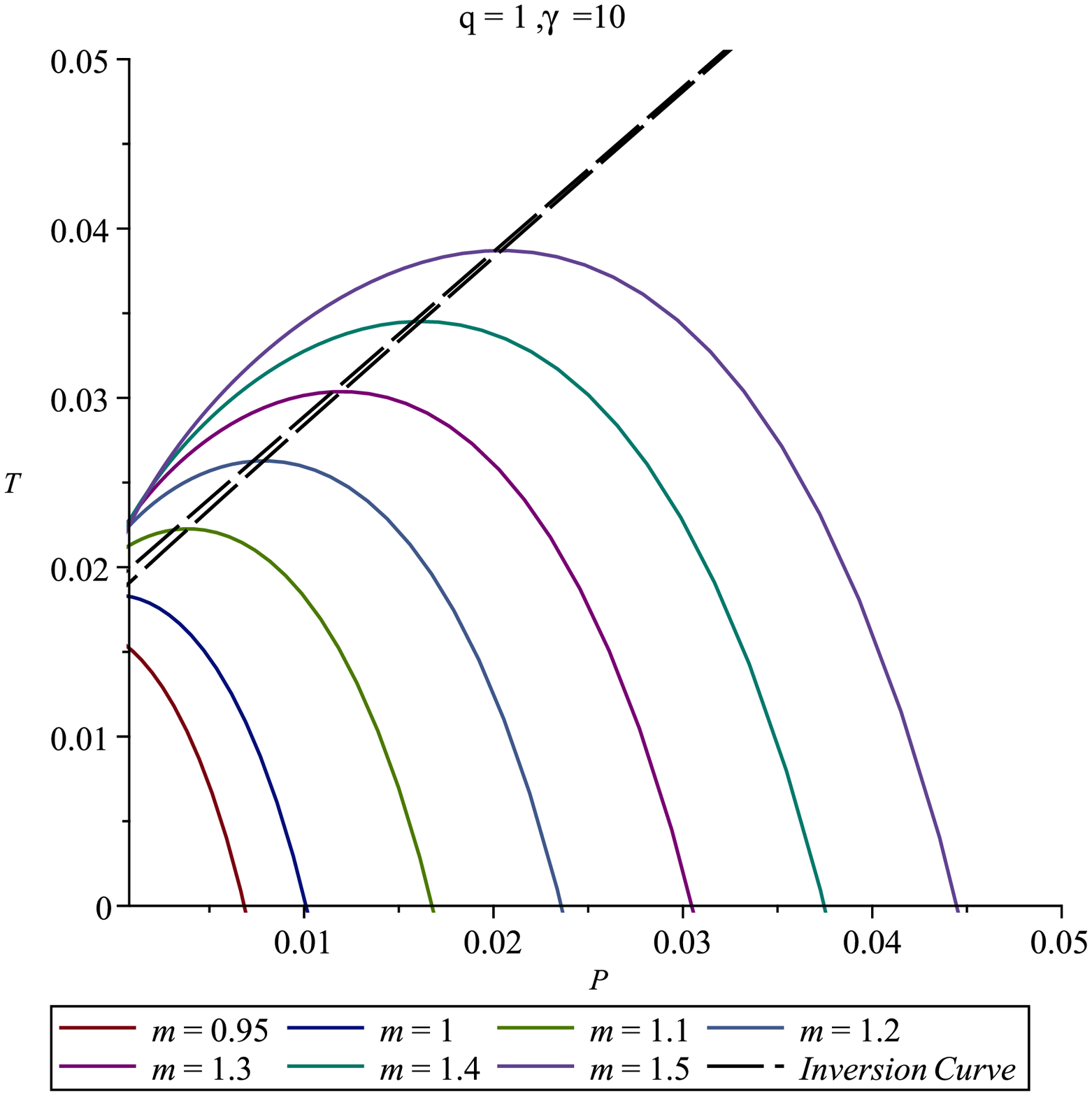}}
\hspace{3mm} \subfigure[{}]{\label{dd}
\includegraphics[width=0.45\textwidth]{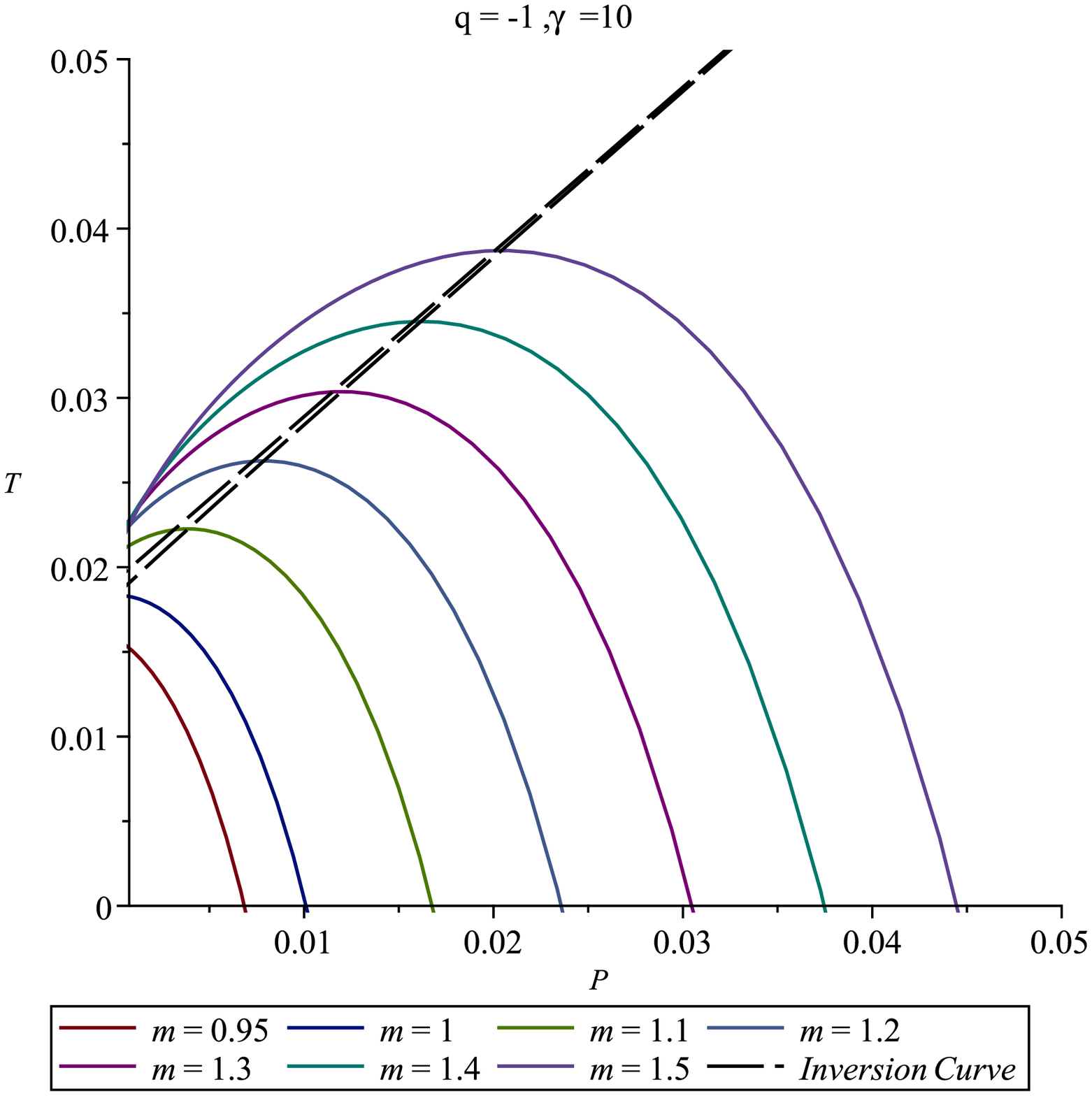}}
\hspace{3mm} \caption{\footnotesize{Intersection of the inversion
curve (the black line) with isenthalpic curves for $q=\pm 1$ and
$\gamma=10$}}
\end{figure}
\begin{figure}[ht]
\hspace{3mm} \subfigure[{}]{\label{ee}
\includegraphics[width=0.45\textwidth]{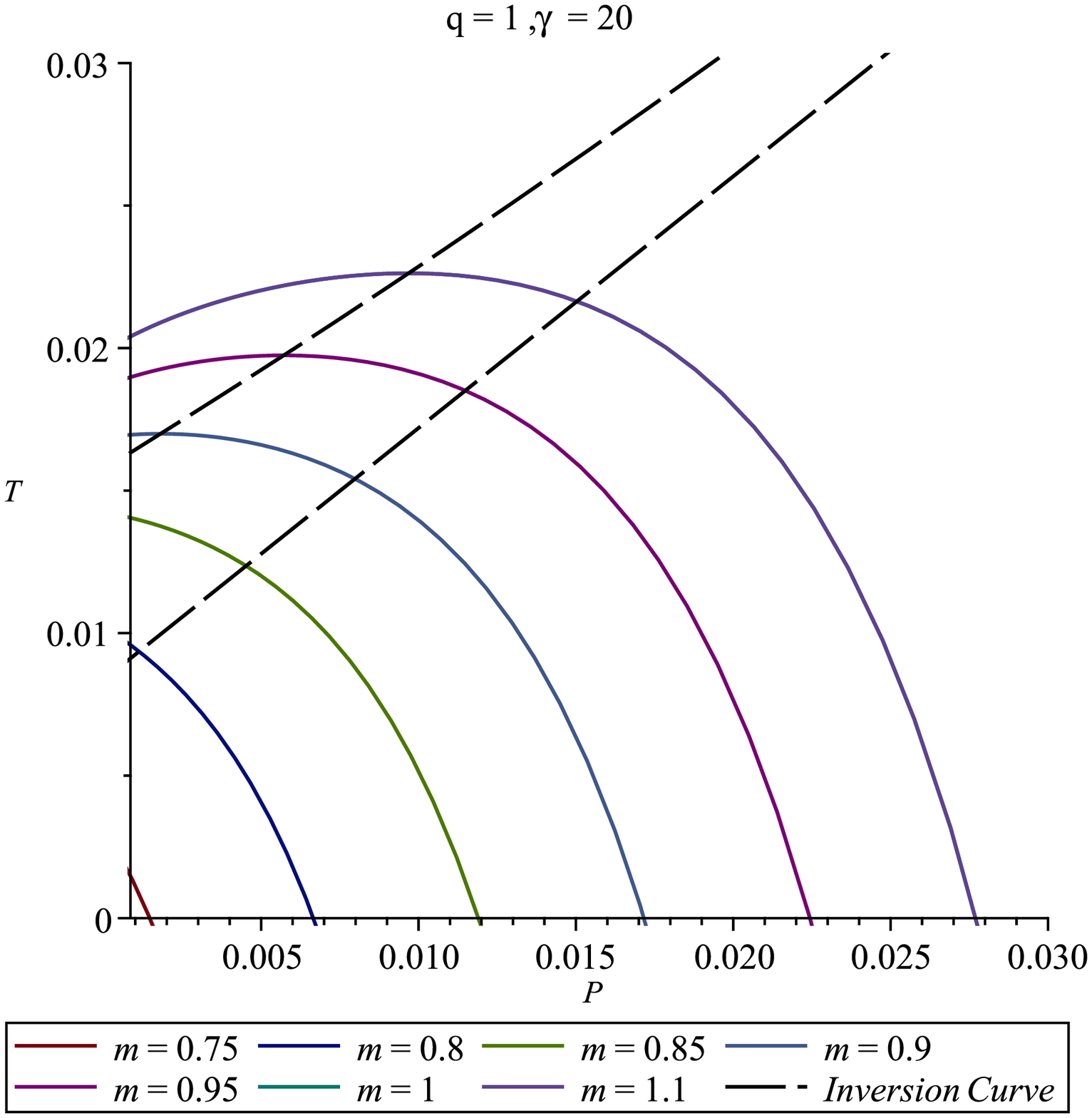}}
\hspace{3mm} \subfigure[{}]{\label{ff}
\includegraphics[width=0.45\textwidth]{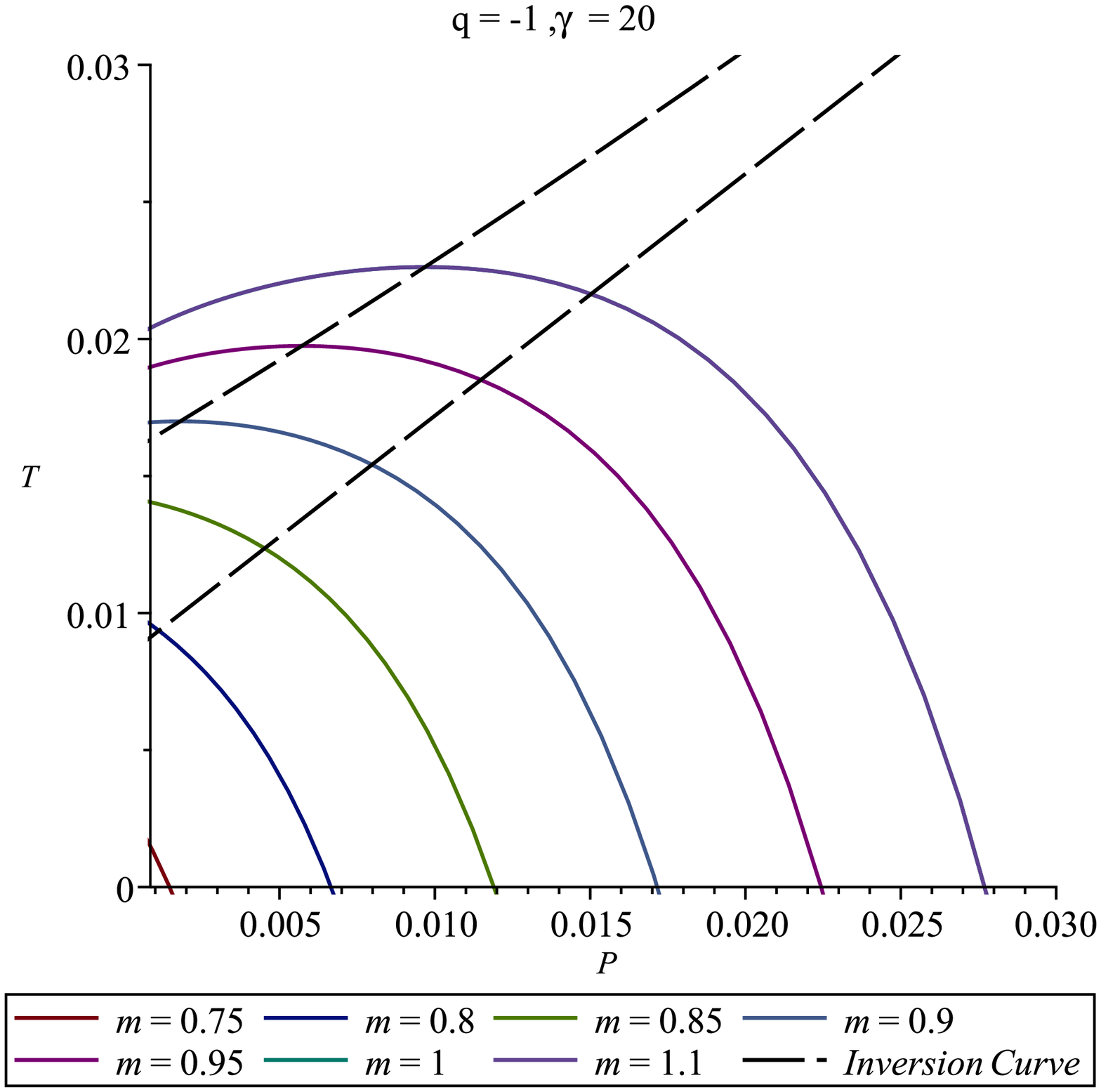}}
\hspace{3mm} \caption{\footnotesize{Intersection of the inversion
curve (the black line) with isenthalpic curves for $q=\pm 1$ and
$\gamma=20$}}
\end{figure}
\begin{figure}[ht]
\hspace{3mm} \subfigure[{}]{\label{aaa}
\includegraphics[width=0.45\textwidth]{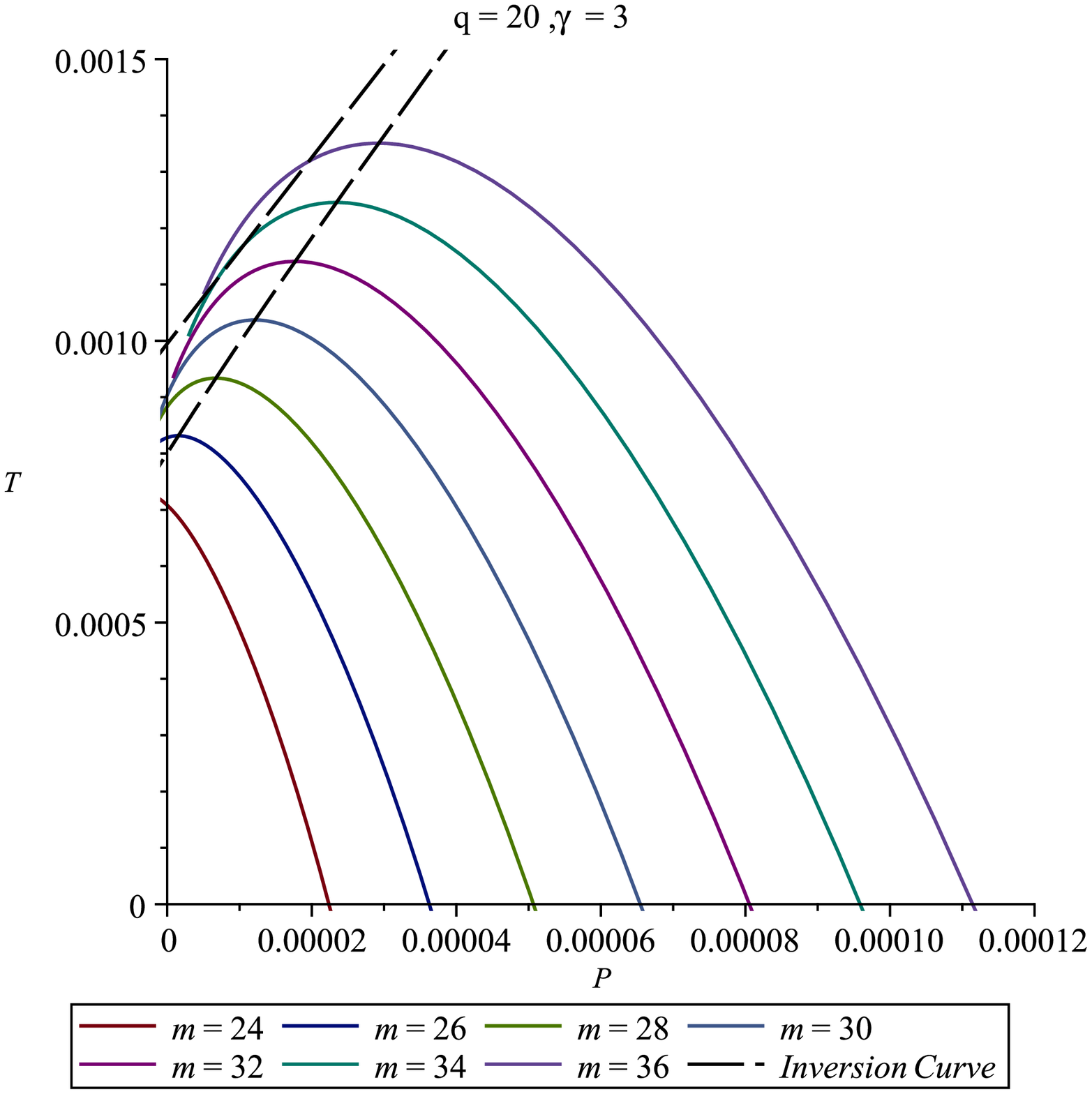}}
\hspace{3mm} \subfigure[{}]{\label{bbb}
\includegraphics[width=0.45\textwidth]{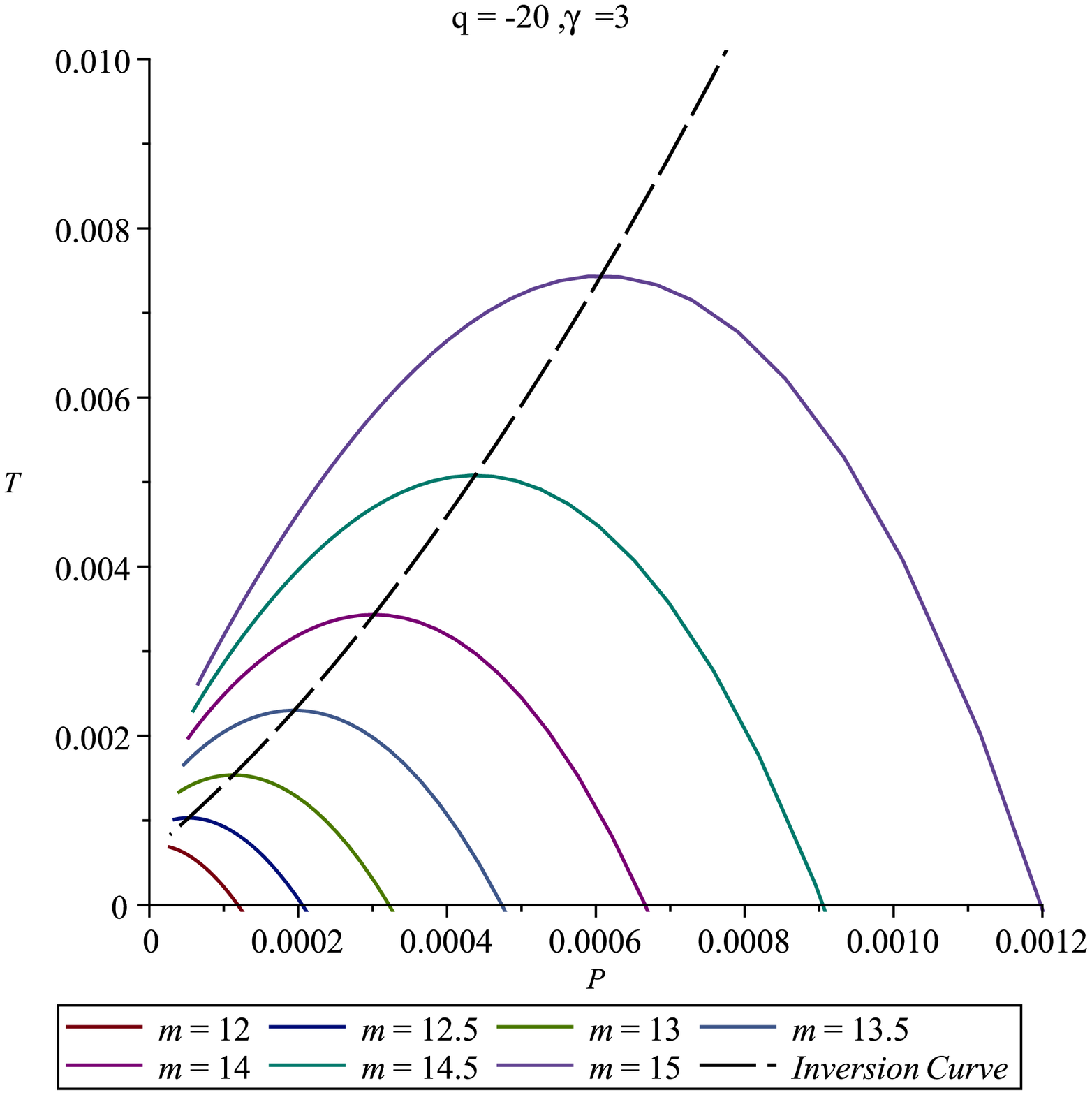}}
\hspace{3mm} \caption{\footnotesize{Intersection of the inversion
curve (the black line) with isenthalpic curves for $q=\pm 20$ and
$\gamma=3$}}
\end{figure}
\begin{figure}[ht]
\hspace{3mm} \subfigure[{}]{\label{ccc}
\includegraphics[width=0.45\textwidth]{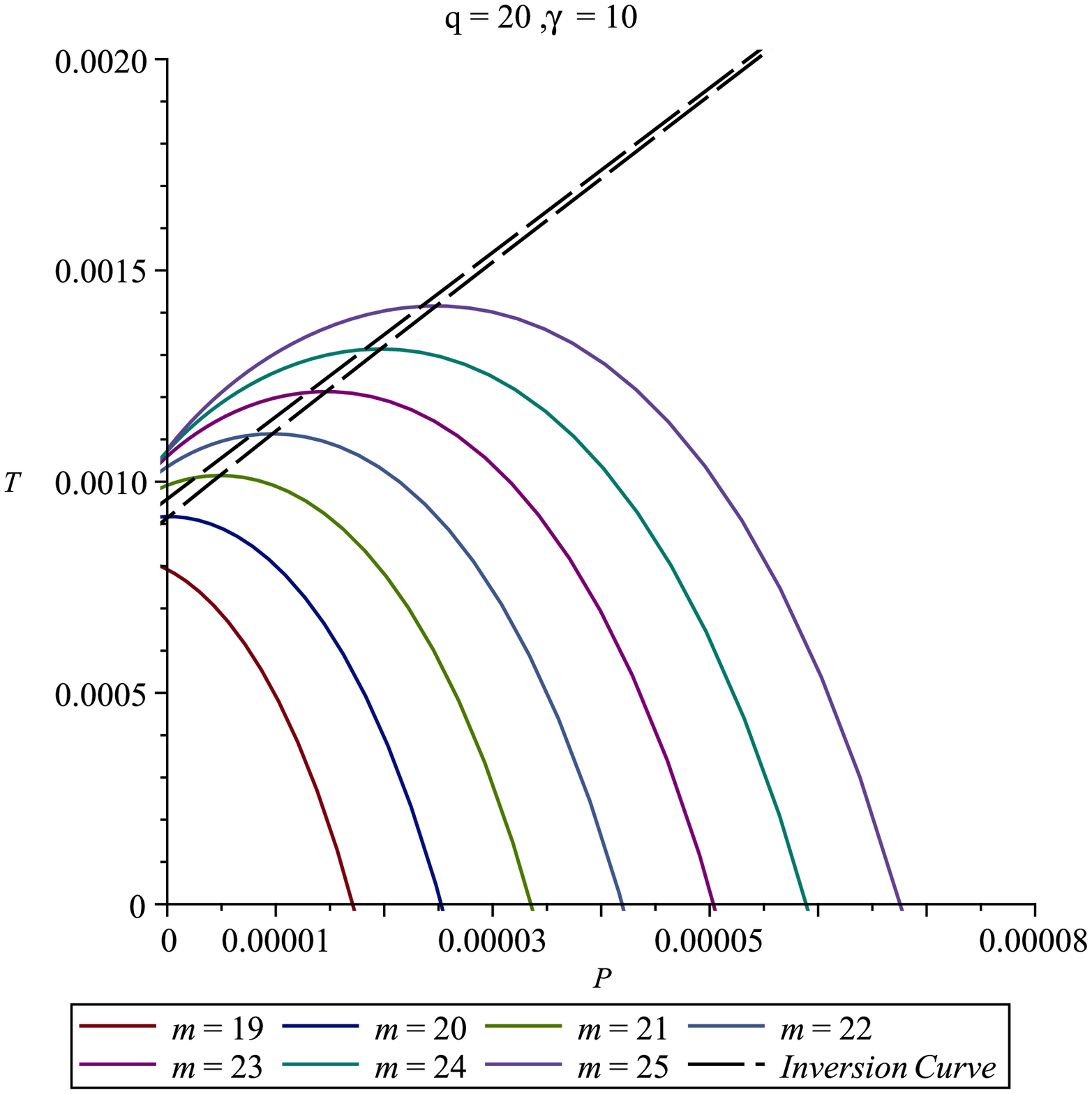}}
\hspace{3mm} \subfigure[{}]{\label{ddd}
\includegraphics[width=0.45\textwidth]{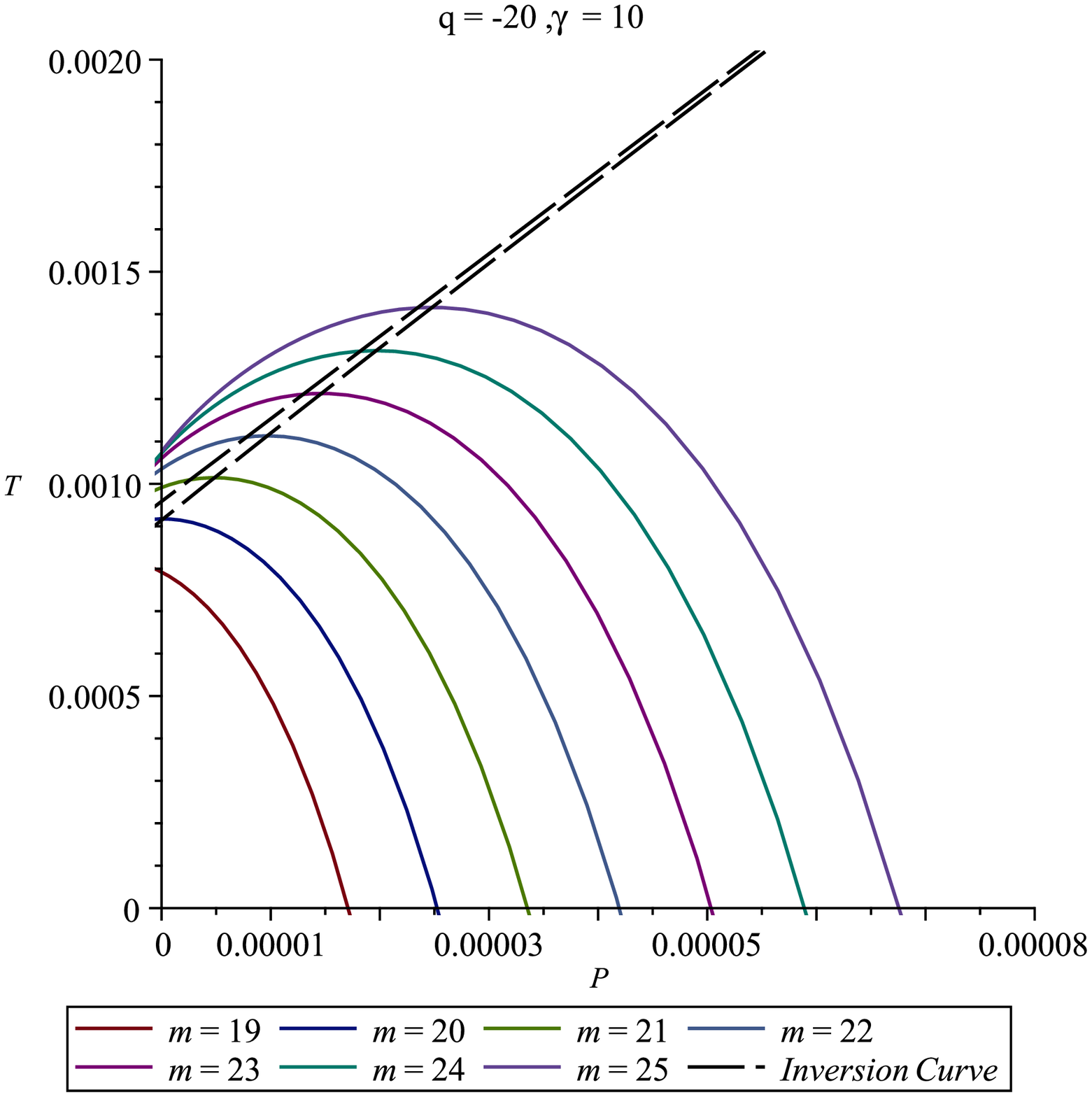}}
\hspace{3mm} \caption{\footnotesize{Intersection of the inversion
curve (the black line) with isenthalpic curves for $q=\pm 20$ and
$\gamma=10$}}
\end{figure}
\end{document}